\PassOptionsToPackage{bookmarks={false}}{hyperref}
\documentclass[sigconf,9pt]{acmart}

\renewcommand\footnotetextcopyrightpermission[1]{} 
\usepackage{amsmath}
\usepackage{times}
\usepackage{soul}
\usepackage{color}
\usepackage{url}
\usepackage{amsthm}
\usepackage{algorithm}
\usepackage{algorithmic}

\newenvironment{myitemize}{\begin{list}{$\bullet$}
{\setlength{\topsep}{1mm}
\setlength{\itemsep}{0.25mm}
\setlength{\parsep}{0.25mm}
\setlength{\itemindent}{0mm}
\setlength{\partopsep}{0mm}
\setlength{\labelwidth}{15mm}
\setlength{\leftmargin}{4mm}}}{\end{list}}

\AtBeginDocument{%
  \providecommand\BibTeX{{%
    \normalfont B\kern-0.5em{\scshape i\kern-0.25em b}\kern-0.8em\TeX}}}

\copyrightyear{2020}
\acmYear{2020}
\setcopyright{acmcopyright}\acmConference[ICCAD '20]{IEEE/ACM International Conference on Computer-Aided Design}{November 2--5, 2020}{Virtual Event, USA}
\acmBooktitle{IEEE/ACM International Conference on Computer-Aided Design (ICCAD '20), November 2--5, 2020, Virtual Event, USA}
\acmPrice{15.00}
\acmDOI{10.1145/3400302.3415676}
\acmISBN{978-1-4503-8026-3/20/11}

\begin{document}


\title{Energy-Efficient Control Adaptation with Safety Guarantees for Learning-Enabled Cyber-Physical Systems}


\author{Yixuan Wang, Chao Huang and Qi Zhu}
\email{yixuanwang2024@u.northwestern.edu, {chao.huang,  qzhu}@northwestern.edu}
\affiliation{%
  \institution{Electrical and Computer Engineering, Northwestern University, USA}
  \postcode{60208}
}

\renewcommand{\shortauthors}{Yixuan, Chao and Qi.}

\begin{abstract}
Neural networks have been increasingly applied for control in learning-enabled cyber-physical systems (LE-CPSs) and demonstrated great promises in improving system performance and efficiency, as well as reducing the need for complex physical models. However, the lack of safety guarantees for such neural network based controllers has significantly impeded their adoption in safety-critical CPSs. In this work, we propose a controller adaptation approach that automatically switches among multiple controllers, including neural network controllers, to guarantee system safety and improve energy efficiency. 
Our approach includes two key components based on formal methods and machine learning. First, we approximate each controller with a Bernstein-polynomial based hybrid system model under bounded disturbance, and compute a safe invariant set for each controller based on its corresponding hybrid system. Intuitively, the invariant set of a controller defines the state space where the system can always remain safe under its control. The union of the controllers' invariants sets then define a safe adaptation space that is larger than (or equal to) that of each controller. 
Second, we develop a deep reinforcement learning method to learn a controller switching strategy for reducing the control/actuation energy cost, while with the help of a safety guard rule, ensuring that the system stays within the safe space.  
Experiments on a linear adaptive cruise control system and a non-linear Van der Pol's oscillator demonstrate the effectiveness of our approach on energy saving and safety enhancement.  
\end{abstract}


\maketitle
\fancyhead[]{}

\section{Introduction}
\label{sec:introduction}

Learning-enabled cyber-physical systems (LE-CPSs)~\cite{xiang2018verification, tuncali2018reasoning, cai2020real, hartsell2019model} often leverage machine learning techniques in their perception of the environment, and increasingly also in the consequent decision making process for planning, navigation, control, etc. In particular, neural network based controllers have been applied to a variety of LE-CPSs, such as building HVAC control~\cite{wei2017deep}, autonomous vehicles ~\cite{lin2018deep}, smart grid ~\cite{lu2018dynamic} and robotics ~\cite{xu2018feedback}, due to their improvement on control performance and efficiency, and the fact that they do not require building a complex physical model of system dynamics.   
However, the uncertainties from the system input and the neural network itself make it quite challenging to ensure the safety of neural-network controlled systems, which has significantly hindered their adoption in safety-critical CPSs~\cite{xiang2018reachability}. 

In this work, we present an approach to leverage multiple controllers (including but not limited to neural network controller) and design an intelligent adaptor for switching among them to enhance both system safety and efficiency. 
At each sampling instant, the adaptor will choose the appropriate controller based on the current system state, and then applies the control input computed by the chosen controller. 
Our approach is motivated by the intuition that for many CPSs, multiple controllers designed based on different methodologies may each have their advantages at different system states. Thanks to the rapid advancement in learning-based control, there are a variety of learning methodologies that can help build neural network controllers for a system~\cite{lillicrap2015continuous, fujimoto2018addressing, haarnoja2018soft}. 
In addition, well-established model-based controllers, such as PID~\cite{PID}, LQR~\cite{LQR} and MPC~\cite{MPC}, have their own advantages and could be complementary to data-driven neural network controllers.
Then, with effective adaptation/switching strategy, multiple such controllers can jointly provide a larger operation space the facilities the improvement of system safety and efficiency.  

With this intuitive motivation, our approach addresses two key technical challenges for \textbf{achieving the guarantee of system safety and the improvement of system energy efficiency}: 
\begin{myitemize}
    \item We develop an invariant-based formal method for analyzing the safe configuration space of each controller to guide the adaptor for making the safe choice. 
    Computing an invariant for classical systems have been extensively explored ~\cite{invlinear, xue2018robust}. However, it still remains an open problem for neural-network controlled systems (NNCSs). To address this challenge, our method provides a general approach to compute the (robust) invariant set for a large variety of controllers, including linear, polynomial, and neural network based ones. First, we approximate each controller with Bernstein polynomials under bounded error, and if the approximation precision is not sufficient, further refine the approximation by partitioning the system state space. Then, using over-approximation, we convert the system with each controller to a hybrid polynomial system under bounded disturbance and compute its (robust) invariant set with semi-definite programming (SDP)~\cite{xue2018robust}.
    After obtaining the invariant of each controller, the adaptor can ensure the system safety by only choosing from the controllers whose invariant set covers the current system state. 
    \item Given the computed invariant sets for the controllers, the second challenge is to intelligently switch among the controllers for reducing the energy consumption while guaranteeing safety. An effective strategy should select the appropriate controller from all safe choices to reduce the \emph{overall} energy. Given the complexity and heterogeneity of multiple controllers, traditional methods based on optimization techniques can hardly handle it. Thus, we develop a deep reinforcement learning (DRL) algorithm that automatically learns the adaptation strategy among safe controllers. At each sampling instant, the adaptor make a choice among the safe controllers based on the current system state, and find the most efficient one for reducing overall energy consumption. This is achieved by a carefully designed reward function in learning and a safety guard rule to discard the rare unsafe choice. 
\end{myitemize}



\smallskip
\noindent
\textbf{\textit{Related work:}}
Our work is related to a rich literature on the safety verification of controlled systems. General safety verification relies on the computation of the reachable set, which contains all possible system states after a finite time for a given initial state set. Existing techniques falls into two main categories: 1) explicitly evaluating the reachable set ~\cite{anai2001reach, tomlin2003computational, asarin2003reachability, huang2019reachnn}, and 2) implicitly considering the reachable set such as barrier certificates  ~\cite{romdlony2016stabilization,prajna2006barrier,yang2016linear,huang2017probabilistic}. The main difference between the invariant set in our approach and the reachable set in the literature is that the invariant set enables infinite-time safety verification while the reachable set provides a finite-time horizon. 
 In~\cite{huang2019reachnn,fan2020reachnn}, Bernstein polynomials are applied in reachable set computation to approximate neural network controllers, but only on a small part of the state space. In contrast, our method applies Bernstein polynomials on the entire space for invariant set computation.

As we develop a DRL-based method with safety guarantees, our approach is related to the research topic of safe reinforcement learning ~\cite{alshiekh2018safe, junges2016safety, kroening2020towards}. The action exploration in RL causes the unsafe state. Thus, one idea is to force the agent to explore within the action set that is a prior known to be safe at a given state ~\cite{garcia2015comprehensive}. Our approach falls into the same idea but with the formally verified safety results. Formal methods are also used in ~\cite{fulton2018safe} for linear adaptive cruise control(ACC) with the tabular Q-learning method. In contrast, our approach mainly targets neural network controllers. 

Our work is also related to ~\cite{huang2020opportunistic}, which also tries to reduce system energy consumption while guaranteeing its safety. In particular, it guarantees the safety by deriving three different levels of safety sets and reduces the energy consumption by skipping the control input. However, that approach cannot be applied to neural network controllers, which is the focus of this work. 

\smallskip
\noindent
In summary, our work makes the following contributions:
\begin{myitemize}
    \item We develop a novel framework for  energy-efficient control with safety guarantees by intelligently switching among multiple controllers (including neural network controllers) for LE-CPSs.  
    \item Our framework guarantees infinite-time system safety, as long as the 
    initial state is within the joint safe configuration space computed through a novel Bernstein polynomial based controller approximation method.   
    \item We develop a new DRL method to learn an adaptation strategy that reduces the overall control energy consumption, while ensuring the system stay within the safe space.
    \item We conduct extensive experiments on a linear ACC system and a non-linear Van der Pol's oscillator system. The results indicate the effectiveness of our approach in enhancing system safety and energy efficiency, when compared with using a single controller.
\end{myitemize}

 The rest of the paper is organized as follows. Section~\ref{sec:problem_formulation} introduces an illustrating example and defines problem formulation. Section~\ref{sec:our_approach} presents our approach. Section~\ref{sec:experiment} shows the experimental results, and Section~\ref{sec:discussion} provides further discussion. Section~\ref{sec:conclusion} concludes the paper. 
 
\section{Problem Formulation}
\label{sec:problem_formulation}

We will start with an illustrating example that helps explain the problems we are trying to solve, and then formally formulate them.

\smallskip
\noindent
\textbf{Illustrating Example [Van der Pol's Oscillator]:} Van der Pol's oscillator ~\cite{barbosa2007analysis} is a 2-dimensional non-linear system whose discrete-time dynamics is given as
\begin{equation} \label{OS_dynamics}
\begin{cases}
x_{1}(t{+}1) =x_{1}(t) + x_{2}(t)\delta\\
x_{2}(t{+}1) =x_{2}(t) + \delta[(1{-}x_{1}^{2}(t))x_{2}(t) {-} x_{1}(t) {+} u(t)] + \omega(t)
\end{cases}
\end{equation}
where $\delta=0.05$ is the sampling period, $u(t)$ is the control input, and $\omega(t)$ is the external disturbance that is uniformly random distributed over $[-0.05, 0.05]$. $(x_{1}, x_{2})$ are the state variables. The safe state space is a box $[-2, 2]*[-2, 2]$. 

 Previous works ~\cite{wei2018general, heydari2014revisiting, xu2018learning, mesbah2016stochastic} have designed neural networks to control the oscillator to the origin point. In this paper, we use two neural network controllers $\kappa_{1}$ and $\kappa_{2}$ for the oscillator that are designed with the DDPG method~\cite{lillicrap2015continuous}, as detailed in Section~\ref{sec:experiment}.

Assume the oscillator is at an initial state $(1, 1)$ within the safe space, we are interested in the following questions. Does the system always stay within the safe box by applying $\kappa_{1}$? 
If not, from what other initial states, the system could be always safe by applying $\kappa_{1}$? Similar questions could be asked for the oscillator with controller $\kappa_{2}$. 
 Then, if we verify that system with the initial state $(1, 1)$ can be safely controlled by either $\kappa_{1}$ or $\kappa_{2}$, which controller should we pick for the overall energy reduction?
 Trying to answer these questions motivates our formal definition of the problems below and our proposed approach. The illustrating example will be used throughout the paper and its solution will be shown in the experiments in Section~\ref{sec:experiment}.
 
\smallskip
\noindent
\textbf{Formulation:} We consider a discrete-time polynomial system: 
\begin{equation} \label{system}
x(t+1) = f(x(t), u(t), \omega(t)), \forall t \geq 0,
\end{equation}
where $x(t) \in \mathbb{R}^{n}$ is the state variable, $u(t) \in \mathbb{R}^{m}$ is the feedback control input variable, $\omega(t) \in \mathbb{R}^{k}$ is a bounded external disturbance, and $f:\mathbb{R}^{n} \times \mathbb{R}^{m} \times \mathbb{R}^{k} \rightarrow \mathbb{R}^{n}$ is a polynomial function.

The safe state space, the constraints on control input, and the external disturbance are given by
\begin{equation} \label{eq:constraint}
x(t) \in X, \quad u(t) \in U, \quad \omega(t) \in \Omega,
\end{equation}
where $X =\{x \in \mathbb{R}^{n} | \bigwedge_{i=1}^{n_{0}} h_{0, i}(x) \leq 0\}$, $ U \in \mathbb{R}^{m}$ and $\Omega = \{\omega \in \mathbb{R}^{k} | \bigwedge_{i=1}^{n_{\omega}} h_{\omega,i}(\omega) \leq 0\}$. $h$ denotes the linear box constraint function. Moreover, We use 1-norm $||u(t)||_{1}$ to denote the control/actuation energy consumption over time step $t$ in this paper.

 The trajectory $\varphi_{x(0)}$ to the system (\ref{system}) starting from an initial state $x(0) \in X$ follows the discrete dynamics denoted by
\begin{displaymath}
\varphi_{x(0)}(t+1) = f(\varphi_{x(0)}(t), u(t), w(t)),
\end{displaymath}
where $\varphi_{x(0)}(0)=x(0)$.
 As stated in Section~\ref{sec:introduction}, we may obtain/design multiple continuous controllers $\kappa_{i}(i=1,2,\cdots, M)$ for such a system, including neural network controllers. Then, the first problem we want to address is the safety verification of the system with each controller $\kappa_{i}$, formulated as the Problem~\ref{P1}. 

\newtheorem{problem}{Problem}
\begin{problem}\label{P1}
Given a dynamical system defined with Equation~\eqref{system} and ~\eqref{eq:constraint} and $M$ continuous controllers $\kappa_{i}(i=1, 2,\cdots, M)$ including neural network controllers, the safety verification problem for the system with each controller $\kappa_{i}$ is to determine whether the controlled trajectory $\varphi_{x(0)}(t) \in X$, $\forall t \geq 0$,  $\forall \omega(t)\in \Omega, \forall x(0) \in X$.   
\end{problem}

With the verification results of the above problem, we then want to design an adaptation strategy $g(x(t)): \mathbb{R}^{n} \rightarrow \{1, \cdots, M\}$ to reduce the overall energy consumption by switching among controllers based on the system state. Here $g$ maps the system state at each time step $t$ to a controller choice. The overall control energy consumption is defined as in Definition~\ref{energy}, and the adaptation optimization problem with safety guarantees is formulated as the Problem~\ref{P2}.

\begin{definition}\label{energy} 
If with infinite-time safety guarantee, the overall control energy consumption of the system in Equation~\eqref{system} as a function of the adaptation strategy $g$ is defined as \footnote{$\kappa_{g}$ is short for $\kappa_{g(x(t))}$ in this paper.} 
\begin{displaymath}
  e(g) = \sum_{t=0}^{+\infty}||\kappa_{_g} (x(t))||_{1}
\end{displaymath}
\end{definition}

\begin{problem} \label{P2}
Given a system defined with Equation~\eqref{system} and ~\eqref{eq:constraint} and multiple continuous controllers $\kappa_{i}(i=1,2,\cdots, M)$ including neural network controllers, and $\forall x(0) \in X$, the problem of optimizing the overall energy consumption with safety guarantee by adaptation strategy function $g$ is formulated as
\begin{displaymath}
\left\{
             \begin{array}{lr}
             \min\limits_{_g} e(g), &  \\
             s.t.\ x(t+1) = f(x(t), \kappa_{_g}(x(t)), w(t)),\forall t \geq 0 & \\
             \varphi_{x(0)}(t) \in X, \forall t \geq 0, \forall \omega \in \Omega&
             \end{array}
\right.
\end{displaymath}
\end{problem}

\section{Energy-efficient Controller Adaptation with Safety Guarantee}
\label{sec:our_approach}

As stated in Section~\ref{sec:introduction}, there are two key aspects of our approach: 1) computing the robust invariant set of each controller to build a joint safe configuration space, and 2) developing a DRL-based method to learn an efficient adaptation strategy within the joint safe configuration space.

For 1), informally, robust invariant set $X_{I}^{i} \subseteq X$ of the controller $\kappa_{i}$ is a set that any controlled trajectory starting from it will never leave it under any possible disturbance within $\Omega$. 
To compute the $X_{I}^{i} (i=1, 2,,\cdots, M)$, we first apply Bernstein polynomials with bounded error to overly approximate each controller via state space partition. This approximation converts each original controlled system such as an NNCS into a hybrid polynomial system with bounded disturbance. We can then obtain the inner-approximation of the $X_{I}^{i}$ with SDP by using existing techniques ~\cite{xue2018robust}. After that, we build the joint safe configuration space as the union of the computed inner-approximations of robust invariant sets, within which the infinite-time safety is guaranteed for the system.
 
For 2), we develop a DRL method to learn an efficient adaptation strategy within the joint safe configuration space, thus guaranteeing the system safety. More specifically, we set a reward function for punishing large control input and unsafe controller choice, so that the DRL agent can learn to reduce the energy consumption while maintaining safety. 
In the rare case that the DRL agent selects an unsafe controller choice, a safety guard rule will discard it and randomly choose a safe controller instead.
 
\begin{figure}
    \centering
    \includegraphics[width=0.9\linewidth]{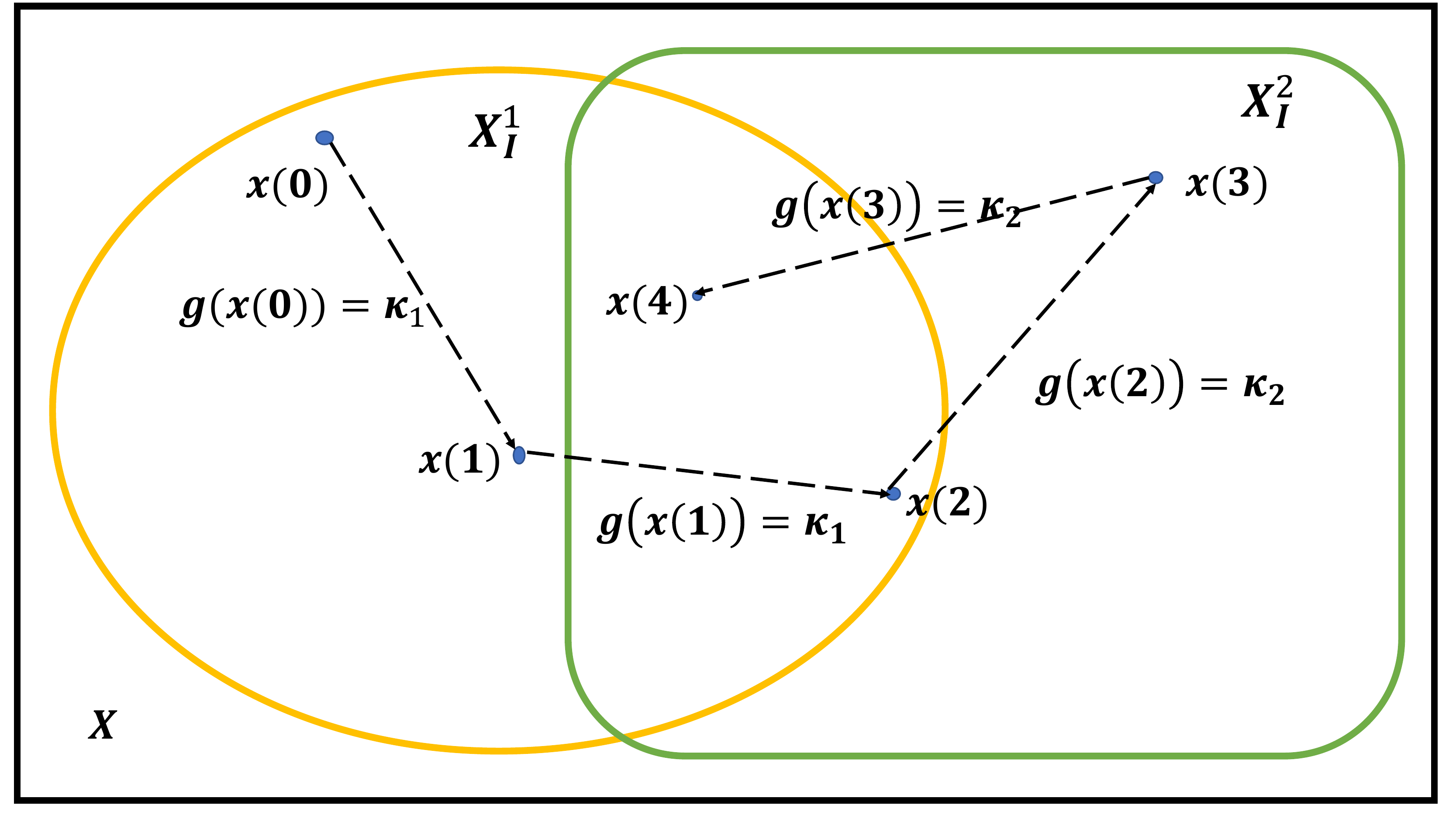}
    \caption{Illustration of the schematic of our approach: Consider the oscillator with two NN controllers $\kappa_{1}$ and $\kappa_{2}$. Here $X$ is the defined safe state space. Assume $X_{I}^{1}, X_{I}^{2}$ are the robust invariant sets for each controller, respectively. The joint safe configuration space is $X_{I}^{1} \cup X_{I}^{2}$. For safety guarantee, when system is at the state  $x(3) \in X_{I}^{2}$, we should choose $\kappa_{2}$. For energy efficiency, when system is at the state $x(2) \in X_{I}^{1} \cap X_{I}^{2}$, where it can be safely controlled by using either $\kappa_{1}$ or $\kappa_{2}$, the adaptor decides to choose controller $\kappa_{2}$ to reduce overall control energy cost.}
    \label{schematic}
\end{figure}

 The schematic of our approach is illustrated in Figure~\ref{schematic}. Its overall framework is described in Algorithm~\ref{alg:Framwork}. 

\begin{algorithm}[htb] 
\caption{Framework of Our Approach.} 
\label{alg:Framwork} 
\begin{algorithmic}[1] 
\REQUIRE 
Multiple controllers $\kappa_{i} (i=1,2,\cdots, M)$ for the system\\
\STATE Compute robust invariant set $X_{I}^{i}$ for each controller $\kappa_{i}$.

\STATE Build the joint safe configuration space as $\cup_{i=1}^{M} X_{I}^{i}$.

\STATE Learn the adaptation strategy $g$ for reducing energy consumption and maintaining system state within $\cup_{i=1}^{M} X_{I}^{i}$ (see Algorithm~\ref{alg:ddqn}).

\STATE Initialization: $t \leftarrow 0$, $x(0) \in  \cup_{i=1}^{M} X_{I}^{i}$.

\WHILE{\textit{true}}
\STATE Read the system state $x(t)$.

\STATE Adaptor $g$ selects controller $\kappa_{_g}$ based on $x(t)$, with safety guard rule applied if needed.

\STATE Actuate the control input $\kappa_{_g}(x(t))$.
\STATE $t \leftarrow t + 1$
\ENDWHILE

\end{algorithmic}
\end{algorithm}

\subsection{Deriving Joint Safe Configuration Space for Safety Guarantee}  

In this section, we show how to compute $X_{I}^{i}$ for the system with $\kappa_{i}$.
We first formally define the concept of robust invariant set $X_{I}^{i}$.

\begin{definition}\label{def1}

Consider a system where the dynamics are defined as Equation~\eqref{system} and the constraint is defined in Equation~\eqref{eq:constraint}. For a controller $\kappa$, $X_{I}$ is called an invariant if
\begin{displaymath}
X_{I} = \{x(0) \ | \ \forall t \geq 0,  \omega(t) \in \Omega, \cdot\ \varphi_{x(0)}(t) \in X_{I}\}.
\end{displaymath}
Moreover, any set that is a subset of the invariant is called an \emph{inner-approximate} invariant.
\end{definition}

Let $X_{I}^{i}$ be the invariant for the $i$-th controller. Then, the joint safe configuration space by multiple controllers can be built as $\cup_{i=1}^{M} X_{I}^{i}$, within which the infinite-time safety is guaranteed for the system.

\begin{proposition}\label{prop1}
(Soundness). For any initial state $x(0) \in \cup_{i=1}^{M} X_{I}^{i}$, the system where  dynamics and constraints are defined in Equation~\eqref{system} and ~\eqref{eq:constraint} with controllers $\kappa_{i} (i=1, 2, \cdots, M)$ is ensured to have infinite-time safety guarantee.
\end{proposition}
\textit{Proof}. Given any initial state $x(0) \in \cup_{i=1}^{M} X_{I}^{i}$, we can at least find one feasible controller $\kappa_{j}$ such that  $x(0) \in X_{I}^{j}$. Then, the system safety is ensured if we always choose $\kappa_{j}$ as the system controller, since as due to Definition \ref{def1}, the controlled trajectory $\varphi_{x(0)}(t) \in X_{I}^{j} \subseteq X, \forall t \geq 0, \forall \omega \in \Omega$.

\newtheorem{remark}{Remark}
\begin{remark}
In general, it is intractable to compute the exact robust invariant set $X_{I}^{i}$ for a nonlinear system~\cite{de2004computation}, especially for neural-network controlled systems. 
Thus in this paper, we compute an inner-approximation of the robust invariant set for the system with each controller, as the inner-approximation maintains the safety guarantee and is more tractable~\cite{de2004computation}. For simplicity, we somewhat abuse the notation for $X_{I}$. \textbf{When we use $X_{I}^{i}$ in the rest of this paper, we point to the inner-approximation of robust invariant set for $\kappa_{i}$.} 
\end{remark}

To compute $X_{I}^{i}$, we first want to approximate controller $\kappa_{i}$ with polynomials under bounded error. This is because neural network controllers are complex and hard to tackle with, while the polynomials are more tractable. This approximation converts the original controlled system such as an NNCS into a polynomial system with bounded disturbance. Prior work~\cite{huang2019reachnn} shows that Bernstein polynomials can be effectively applied to approximate any continuous controller. However, a single polynomial approximation may have to use a very high degree to achieve certain precision, while the computation complexity of $X_{I}^{i}$ increases drastically as the degree increases. Also, the error reduction by this measure is often limited in practice, resulting in an inner-approximation that is too conservative. Thus, following the idea of interpolation, we propose a partition approach to achieve more precise approximation using polynomials with a much lower degree. With such partition approximation, the original controlled system is converted into a hybrid system with low degrees on each subsystem. We can then obtain the inner-approximation of the robust invariant set for such a hybrid system by using SDP. We detail each of these steps in the next. 

\subsubsection{Single Bernstein Polynomial with Bounded Error for Controller Approximation}

We first introduce the concept of Bernstein polynomial.  Let $d=(d_{1}, \cdots, d_{n}) \in \mathbb{R}^{n}$ and $\kappa_{i}$ be a continuous controller of the system over state variables $x=(x_{1},\cdots,x_{n}) \in X$. The polynomials related to controller $\kappa_{i}$
\begin{displaymath}
B_{\kappa_{i}, d}(x) = \sum_{\substack{0\leq a_{j} \leq d_{j} \\ j=\{1, 2,\cdots, n\}}} \kappa_{i} \left(\frac{a_{1}}{d_{1}},\cdots,\frac{a_{n}}{d_{n}} \right) \prod_{j=1}^{n} \left (\tbinom{d_{j}}{a_{j}}x_{j}^{a_{j}}(1-x_{j})^{d_{j}-a_{j}} \right)
\end{displaymath}
are called Bernstein polynomials of $\kappa_{i}$ under degree $d$. 

 To obtain the inner-approximation $X_{I}^{i}$ for the system with controller $\kappa_{i}$, we first overly approximate $\kappa_{i}$ by a single Bernstein polynomial with bounded error in Equation~\eqref{BP} on the safe state space $X$, similar as in~\cite{huang2019reachnn}, 
 \begin{equation}\label{BP}
  \kappa_{i}(x) \in B_{\kappa_{i}, d}(x) + [-\hat{\epsilon}, \hat{\epsilon}], \forall x \in X,
\end{equation}
where $\hat\epsilon$ is the  approximation error bound. Since the controllers in this paper are all considered as continuous functions, according to~\cite{de1959stone}, we can always ensure that such approximation exists. 
 
This approximation converts the system with $\kappa_{i}$ into a polynomial system. The disturbance for the converted system is the  Minkowski sum $\bigoplus$ of external disturbance and approximation error. Now, the system with controller $\kappa_{i}$ is approximated as
\begin{displaymath}
  x(t+1) = f(x(t), B_{\kappa_{i},d}(x(t)),  \hat{\omega}(t)), t \geq 0,
\end{displaymath}
with $\hat{\omega}(t) = $  $\omega(t)\bigoplus \hat{\epsilon}$

However, this single Bernstein polynomial approximation is not sufficient for all encountered neural network controllers in our experiments. 
Recall the oscillator example with the neural network controller $\kappa_{2}$ (details in Section~\ref{sec:experiment}), a single Bernstein polynomial with a low degree, e.g., $d = 3$, for the approximation introduces a large error bound~\footnote{$d=3$ actually means $d=(3, 3)$, representing that the highest polynomial degree for the oscillator state $(x_{1}, x_{2})$ is $(3, 3)$. The same applies to $d=5, 7$.}, as shown in Table~\ref{error_bound}. With such large error bound, we just get an empty set for $X_{I}^{2}$ by SDP.  To reduce the error bound, a simple way is to increase the degree, e.g., set $d=5$ or $7$ for Bernstein polynomial approximation. However, the reduction is limited in practice, as shown in Table~\ref{error_bound}. Moreover, increasing the approximation degree converts the system into a higher order polynomial system, resulting in drastically-increasing computation complexity for $X_{I}^{2}$. Thus, we propose a partition approximation method with low-degree polynomials to reduce the error bound. 
\begin{table}
  \caption{ Error bound by different approximation methods for the oscillator's neural network controller $\kappa_{2}$. The control input space is normalized into interval [-1, 1]. Note that the partition approximation achieves the smallest bound. Simply increasing the degree will reduce the error bound but has limited effect.}
  \label{error_bound}
  \begin{tabular}{cccl}
    \toprule
      3-Partition (d=3) & Single (d=3) & Single (d=5) & Single (d=7) \\
    \midrule
      \textbf{0.102} & 0.27 & 0.169 & 0.163 \\
  \bottomrule
\end{tabular}
\end{table}

\subsubsection{Partition Approximation}

We first partition $X$ into $P$ boxes with each box named as $X^{p}$, for $p= (1, 2, \cdots, P)$:
\begin{displaymath} 
X^{p_{1}} \cap X^{p_{2}} = \emptyset, if \  p_{1} \neq p_{2} \ and \ \cup_{p=1}^{P} X^{p} = X,
\end{displaymath}
where $p_{1}, p_{2} \in \{1, 2,\cdots, P\}$. Now each box $X^{p}$ has its own state constraints, defined as
$X^{p} = \{x \in \mathbb{R}^{n} | \bigwedge_{i=1}^{n_{p}} h_{p,i}(x) \leq 0 \}$, where $h$ denotes the linear box constraint function. 

Then, on each box $X^{p}$, a Bernstein polynomial $B_{\kappa_{i},d}^{p}$ is applied for approximation, reducing the overall approximation error bound  $\hat{\epsilon} = \max(\hat{\epsilon}^{p})$, where $\hat{\epsilon}^{p}$ is the error bound on box $X^{p}$ as
\begin{displaymath}
 \kappa_{i}(x) \in B_{\kappa_{i}, d}^{p}(x) +  [-\hat{\epsilon}^{p}, \hat{\epsilon}^{p}],\ \forall x \in X^{p}.
\end{displaymath}

With such partition, the system with each controller can now be converted into a hybrid polynomial system. Each partition now acts as a subsystem with Bernstein polynomial control input on it. For this hybrid system, the new bounded disturbance is the Minkowski sum $\bigoplus$  of external disturbance $\omega$ and overall approximation error bound $\max(\hat{\epsilon}^{p})$. Such a hybrid system can be expressed as
\begin{displaymath}
x(t+1) = f(x(t), \hat{u}(t), \hat{\omega}(t)), t >= 0,
\end{displaymath}
where $\hat{u}(t)$ and  $\hat{\omega}(t)$ are
\begin{equation} \label{hydisturbance}
\hat{u}(t) = \sum_{p=1}^{P} \mathbf{1}_{X^{p}} \cdot B_{\kappa_{i},d}^{p}(x(t)), \quad
  \hat{\omega}(t) = \omega(t)\bigoplus \max (\hat{\epsilon}^{p}),
\end{equation}
where $\mathbf{1}_{X^{p}}$ is an indicator function, $p= (1,2,\cdots, P)$.

When we use the partition approach to approximate the $\kappa_{2}$ of the oscillator with $d=3$, we achieve the smallest error bound, when compared with $d=3, 5, 7$ under the non-partitioned single-polynomial approximation. This is shown in Table~\ref{error_bound}.

\begin{remark}
For polynomial controller $\kappa_{i}$ with degree $d_{0}$, if we choose Bernstein polynomial $B_{\kappa_{i},d_{0}}$ also with degree $d_{0}$, then the approximation error $\hat{\epsilon}=0$. For the feed-forward neural network controller,
the partition approximation greatly reduces $\hat{\epsilon}$ in practice, compared to single-polynomial approximations.
\end{remark}

Next, the inner-approximation of the robust invariant set of such a converted hybrid system is computed. 

\subsubsection{Inner-approximation of Robust Invariant Set}\label{inner_app}

Each converted hybrid system has constraints defined as Definition~\ref{SD}. 

\begin{definition}\label{SD}
Each converted hybrid polynomial system is subject to state constraints on each partition $X^{p}$, the entire safe space $X$ and the disturbance $\hat{\Omega}$ ($\hat{\omega}$ defined in Equation~\eqref{hydisturbance}), which can be expressed as the following sets.
\begin{displaymath}
\begin{cases}
X =\{x \in \mathbb{R}^{n}\ | \bigwedge_{i=1}^{n_{0}} h_{0, i}(x) \leq 0\}\\
 X^{p} = \{x \in \mathbb{R}^{n}\ | \bigwedge_{i=1}^{n_{p}} h_{p,i}(x) \leq 0 \}\\
 \hat{\Omega} = \{\hat{\omega} \in \mathbb{R}^{k}\ | \bigwedge_{i=1}^{n_{\hat{\omega}}} h_{\hat{\omega},i}(\hat{\omega}) \leq 0\}
\end{cases}
\end{displaymath}
where $p=(1, 2, \cdots, P)$, and  $h$ denotes the linear box constraint. 
\end{definition}

Then, following the method in ~\cite{xue2018robust}, the inner-approximation of robust invariant set for such a hybrid system can be obtained by solving an SDP. First, we compute the one-step reachable set $R(X)$ as the states reachable from the $X$ within one-step computation, i.e.,
\begin{displaymath}
  R(X) \coloneqq \{x \ | \ x = f(x,  \hat{u}, \hat{\omega}), x \in X, \hat{\omega} \in \hat{\Omega}\} \cup X.
\end{displaymath}
Then, we define a continuous function $v(x): \mathbb{R}^{n} \rightarrow \mathbb{R}$. When $v(x)$ is constrained to the polynomial type and the system state is constrained in a ball $B$ with $H$ as a constant
\begin{displaymath}
  B = \{x\ |\ ||x||_{2} - H \leq 0\},
\end{displaymath}
such that $R(X) \subseteq B$. Then, according to~\cite{xue2018robust}, the inner-approximation of the robust invariant set as $\{x \in B \ | \ v(x) \leq 0\}$ can be obtained by solving an SDP optimization problem
\begin{displaymath}
\begin{cases}
\min\limits_{v,\ s_{p, l_{1}}^{X^{p}}, \ s_{l_{2}}^{\hat{\Omega}},\ s_{p},\ s_{1, j}^{'}} c \cdot w\\ 
v(x) - v(f(x, \hat{u}, \hat{\omega})) + \sum_{l_{1}=1}^{n_{p}}s_{p, l_{1}}^{X^{p}}h_{i, l_{1}}(x) +\\ \qquad \sum_{l_{2}=1}^{n_{\hat{\omega}}}s_{l_{2}}^{\hat{\Omega}}h_{\hat{\omega},l_{2}}(\hat{\omega})-s_{p}h(x) \in SOS(x, \hat{\omega}),\\
(1+h_{0,j}^{2})v(x)-h_{0, j}(x) -s_{1, j}^{'}h(x) \in SOS(x),
\end{cases}
\end{displaymath}
where $c \cdot w = \int_{B}v(x)dx$, $c$ is the unknown coefficient vector in $v(x)$, and $w$ is the vector of the integration for each monomial in $v(x)$ over $B$.  $s_{p, l_{1}}^{X^{p}}, \ s_{l_{2}}^{\hat{\Omega}},\ s_{p},\ s_{1, j}^{'}$ are the sum-of-squares$(SOS)$ polynomials, where $p=(1, 2, \cdots, P), l_{1} = (1, 2, \cdots, n_{p}), l_{2} = (1, 2, \cdots, n_{\hat{\omega}})$ and $j=(1, 2, \cdots, n_{0})$. $s_{p, l_{1}}^{X^{p}}, \ s_{l_{2}}^{\hat{\Omega}},\ s_{p} \in SOS(x, \hat{\omega})$ and $s_{1, j}^{'} \in SOS(x)$.  

\begin{figure}
    \centering
    \includegraphics[width=\linewidth]{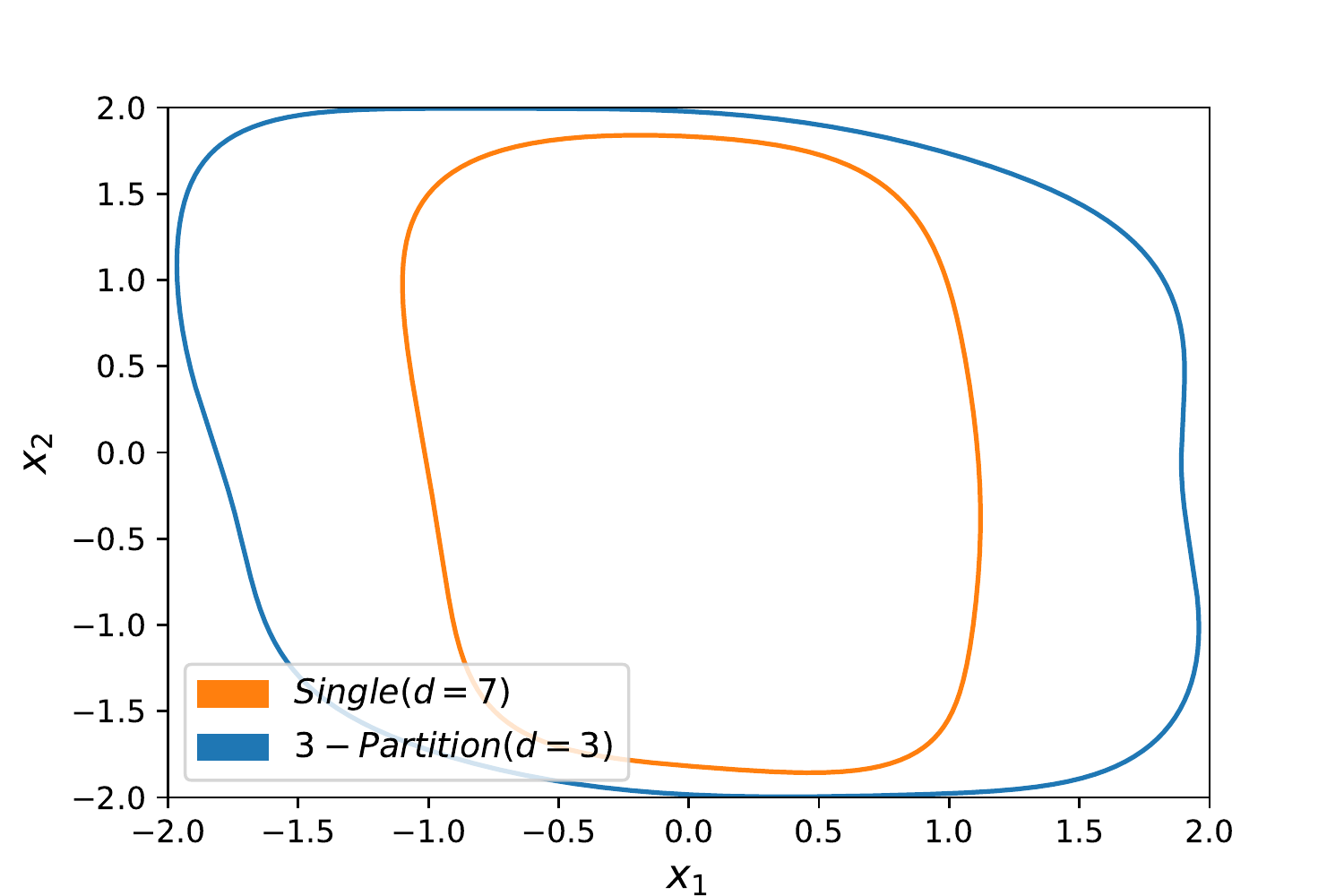}
    \caption{ $X_{I}^{2}$ of oscillator with $\kappa_{2}$ by SDP for different approximation methods. For $Single (d=3)$, the SDP returns an empty set due to its large error bound. For $Single (d=7)$, we obtain a non-empty inner-approximation but it is much more conservative/inaccurate than the $3-Partition$ method, where 3 polynomials are used with the partition approximation. 
    Moreover, it took about 2 hours to compute $X_{I}^{2}$ by $3-Partition$ and 41 hours by $Single(d=7)$ with Mosek 8.0 and Matlab 2015.}
    \label{fig:approximation_comp}
\end{figure}

\begin{figure}
    \centering
    \includegraphics[width=\linewidth]{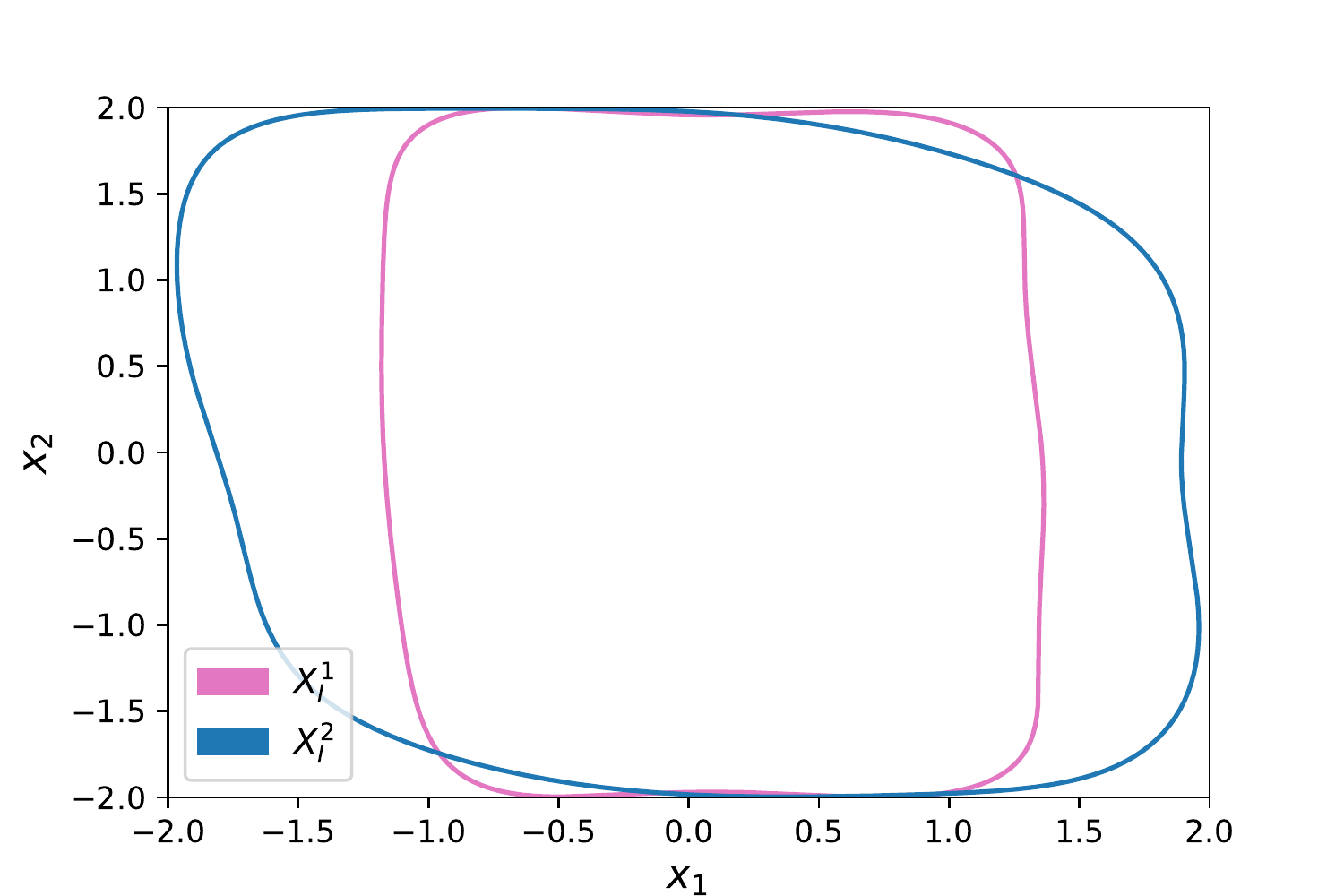}
    \caption{Inner-approximation of the robust invariant sets $X_{I}^{1}, X_{I}^{2}$ for oscillator controlled by the DDPG controllers $\kappa_{1}, \kappa_{2}$. The joint safe configuration space is $X_{I}^{1} \cup X_{I}^{2}$. The controller adaptation learned by DRL will try to reduce the overall energy consumption by intelligently switching $\kappa_{1}$ and $\kappa_{2}$ while maintaining 
   the safety.}
    \label{fig:OS_inv}. 
\end{figure}

\smallskip
\noindent
\textbf{Safe Controller for the Illustrating Example:} Recall the illustrating example. By solving the above SDP problem, we obtain $X_{I}^{2}$ for the controller $\kappa_{2}$ in the oscillator example with different approximation methods. The proposed partition approximation achieves better result than the single-polynomial ones, as shown in Figure~\ref{fig:approximation_comp}. Thus, we use it to obtain 
$X_{I}^{1}$ and $X_{I}^{2}$, as shown in Figure~\ref{fig:OS_inv}.
 It is easy to check that state $(1, 1)$  belongs to the invariant intersection in Figure~\ref{fig:OS_inv}, thus guaranteeing the safety by either $\kappa_{1}$ or $\kappa_{2}$. 

\smallskip
\noindent
Once the joint safe configuration space is derived, we can develop a DRL method to learn an energy-saving adaptation strategy with the safety guarantees, as introduced next.

\subsection{DRL-based Control Adaptation}

Within the safe configuration space $\mathcal{S}=\cup_{i=1}^{M} X_{I}^{i}$, we develop a Double DQN algorithm~\cite{van2016deep} to learn an energy-efficient adaptation strategy with safety guarantees. The learning process can be formulated as a Markov decision process (MDP) with a tuple ($\mathcal{S, A, P, R, \gamma}$). $\mathcal{S}$ represents the state space of MDP. $\mathcal{A}$ is the action space. $\mathcal{P}$ is the state transition probability, mapping the function $\mathcal{S} \times \mathcal{A} \rightarrow \mathcal{S}$. $\gamma$ is the discounted factor, and $\mathcal{R}$ is the reward function encoding the desired goal of the reinforcement learning agent. More specifically, they are formulated as follows. 

\smallskip
\noindent
\textbf{State:} To ensure that the adaptation guarantees safety, the state space $\mathcal{S}$ here is defined as the joint safe configuration space. Moreover, the state of the Double DQN agent is the system state $x(t)$.

\smallskip
\noindent
\textbf{Action:} We define the action space as the discrete space $\mathcal{A}=\{1,\cdots, M\}$. At time $t$, $a(t) \in \mathcal{A}$ means that the Double DQN agent chooses controller $\kappa_{a(t)}$ for controlling the system.

\smallskip
\noindent
\textbf{Reward Function:}
Reward design encodes the desired goals for the agent. First, we set a punishment for the energy cost as $-||u(t)||_{1}$ for the time step $t$. In order to maximize the cumulative reward, the agent needs to learn to avoid large control input. Moreover, the agent needs to set a punishment for choosing any unsafe controller, i.e., choosing controller $\kappa_{a(t)}$ while  $x(t) \notin  X_{I}^{a(t)}$ (note that $x(t) \in \cup_{i=1}^{M} X_{I}^{i}$, which means a safe choice does exist), so that it can learn to avoid such choice. 
With these two considerations, we design the reward function as
\begin{equation}\label{reward}
r(x(t), a(t), x(t+1))=
\begin{cases}
C-\lambda||u(t)||_{1}& \text{Otherwise,}\\
R_{pub}& \text{$if \ x(t) \notin X_{I}^{a(t)}$},
\end{cases}
\end{equation}
where $C$ is a positive constant, $\lambda$ is the weight for the punishment of energy cost $-||u(t)||_{1}$, $R_{pub}$ is a negative constant that punishes the agent for choosing any potential unsafe controller. The reward is -100 when the state is controlled out of the safe space in training. Note that -100 is applied at most once during a training epoch, as the epoch would end after that. 

We develop the Double DQN algorithm to learn an efficient and safe adaptation strategy based on the MDP specified above. The details of the learning process is shown in Algorithm~\ref{alg:ddqn}.

\begin{algorithm}[htb] 
\caption{Double DQN for Learning Adaptation Strategy} 
\label{alg:ddqn} 
\begin{algorithmic}[1] 
\REQUIRE 
Joint safe configuration space $\cup_{i=1}^{M}X_{I}^{i}$\\

\STATE Initialize replay memory $D$, $Q$ network with parameters $\theta$, target network $\hat{Q}$ with parameters $\hat{\theta}$, and update period $C_{0}$.

\FOR{$epoch = 0, \ldots, N$}
\STATE Randomly initialize state $ x(0) \in \cup_{i=1}^{M}X_{I}^{i}$.\

\FOR {$t = 0, \ldots, T$} 

\STATE $a(t) = \epsilon-greedy(Q(x(t)), \epsilon)$.
\IF{$x(t)  \notin X_{I}^{a(t)}$} 
\STATE Update reward punishment $R_{pub}$ and break.
\ENDIF
\STATE Switch to controller $\kappa_{a(t)}$; $x(t)$ evolves to $x(t+1)$; receive reward $r(t)$; store tuple $(x(t), a(t), x(t+1), r(t))$ into $D$.
\STATE Sample mini-batch from $D$; compute TD error ~\cite{sutton1988learning}.
\STATE Apply gradient descent to $Q$.
\STATE Update $\hat{\theta}=\theta$ every $C_{0}$ steps.
\ENDFOR
\ENDFOR
\RETURN $Q$ forwarding function as the adaptation strategy $g$.
\end{algorithmic}
\end{algorithm}

\smallskip
\noindent
\textbf{Safety Guard Rule:} Although we have defined a punishment for any unsafe choice, the Double DQN agent may still occasionally choose unsafe controllers due to the trial-and-error nature of reinforcement learning. In those rare cases, we set a safety guard rule for ensuring system safety. Specifically, if the agent chooses an unsafe controller, the safety guard will discard it and randomly choose a safe one. Note that as long as the system initial state belongs to the joint safe configuration space $\mathcal{S}$, such safe choice always exists.

\smallskip
\noindent
\textbf{Energy-saving Controller for the Illustrating Example:} In this example, the learned Double DQN agent chooses controller $\kappa_{1}$ for the system at the initial state $(1, 1)$, later switches between $\kappa_{1}$ and $\kappa_{2}$, and keeps using $\kappa_{1}$ after around 20 steps as the state is approaching the origin point.  

\section{Experimental Results}
\label{sec:experiment}

Experiments on the illustrating Van der Pol's oscillator example and an adaptive cruise control (ACC) system, a common safety-critical system, are conducted to evaluate the effectiveness of our approach. 

\subsection{Van der Pol's Oscillator}

The Van der Pol's oscillator system is defined in Equation~\eqref{OS_dynamics} in Section~\ref{sec:problem_formulation}. As stated before, we train two controllers by the DDPG method with different reward designs, and name them $\kappa_{1}$ and $\kappa_{2}$. The reward for the DDPG learning can be expressed as (note that this is for learning the underlying controllers $\kappa_{1}$ and $\kappa_{2}$, and different from the Double DQN learning for controller adaptation in Equation~\eqref{reward}):
\begin{displaymath}
r = 10 - \lambda_{1}(|x_{1}| + |x_{2}|) - \lambda_{2}(|u| + |u-u^{'}|),
\end{displaymath}
where 10 is the reward for each safely-controlled step, $\lambda_{1}, \lambda_{2} \geq 0$ are weights for state and control input penalty, respectively, and $u^{'}$ is the control input of previous step. For controller $\kappa_{1}$, both $\lambda_{1} $ and $\lambda_{2}$ are set to 1. For $\kappa_{2}$, $\lambda_{1} $ and $\lambda_{2}$ are set to 5 and 0.2, respectively.

To compute the robust invariant sets, both controllers need to be approximated by Bernstein polynomials with bounded errors via partitioning. Each inner-approximation of the robust invariant set is obtained, as shown in Figure~\ref{fig:OS_inv}.
Then the Double DQN is applied to learn an adaptation strategy between $\kappa_{1}$ and $\kappa_{2}$. The $C$ in the reward Equation~\eqref{reward} is 2,  $\lambda$ is 1, and $R_{pub}$ is -20. The hyper-parameters in Algorithm ~\ref{alg:ddqn} is set as follows: the size of the replay buffer $D$ is 5000, $\gamma$ is 0.99, $C_{0}$ is 100, and the learning rate is 1e-4.  

We set three baselines: using $\kappa_{1}$ only, using $\kappa_{2}$ only, and random adaptation. We conduct 500 test cases by randomly picking 500 initial states within $X_{I}^{1} \cup X_{I}^{2}$, and run all the methods from the same initial state for 200 control steps for each case. 

\begin{table}[h]
  \caption{Comparison of results for the oscillator experiment.}
  \label{OS_result}
  \begin{tabular}{ccccl}
    \toprule
     & Ours & $\kappa_{1}$ only & $\kappa_{2}$ only & Random \\
    \midrule
    Safe control rate & \textbf{100 \%} & 86.4 \% & 95.6 \% & 92 \% \\
    Energy cost & \textbf{127.8} & 130.1 & 164.1 &  383.8 \\
  \bottomrule
\end{tabular}
\end{table}

\smallskip
\noindent
\textbf{Comparison among Different Methods:} 
We compare the average system safety rate and energy cost among different methods, and show them in Table~\ref{OS_result}. Our approach formally guarantees $100\%$ safety as the initial state is within $X_{I}^{1} \cup X_{I}^{2}$, while the other methods all have significant number of unsafe cases. Note that the three baselines do not employ the safety guard rule, since they do not have the capability to compute the safe invariant sets. However, for our approach, even without the safety guard rule, our system is safe for more than $99.6 \%$ of the cases, which shows the effectiveness of Double DQN for switching among controllers.
Moreover, our approach also provides the lowest energy cost, which demonstrates that the reward function design in our Double DQN is effective for overall energy saving.



\subsection{Adaptive Cruise Control}
We also conducted experiments on an ACC system. We consider two vehicles in the system. The front vehicle is running with a  velocity $v_{f}$, while the following/ego vehicle brakes or accelerates according to the control design. 
Overall, the system dynamics is
\begin{displaymath}
\begin{cases}
s(t+1) = s(t) - (v(t) - v_{f}(t))\delta,\\
v(t+1) = v(t) - (kv(t) - u(t)) \delta,
\end{cases}
\end{displaymath}
where $s$ represents the distance between vehicles, $v$ is the velocity of the ego vehicle, $u$ is the control input, $\delta=0.1$ is the sampling period, and $k=0.2$ is the velocity resistance. $v_{f}=40+w$, where $w$ is uniformly random distributed over $[-4, 4]$. The definition of the safe set $X$ over state variable $(s, v)$ is
\begin{displaymath}
X \coloneqq \{(s, v) \ | \ s \in [120, 180], \ v \in [25, 55] \}.
\end{displaymath}

Here we want this ACC system to be controlled stably to the equilibrium state $(150, 40)$. To this end, we design two different controllers -- one is a Linear-Quadratic Regulator (LQR) controller $\kappa_{1}$, and the other is a neural network controller $\kappa_{2}$ obtained by the DDPG method. The LQR's parameters representing the weights for state and control input are set to 2 and 0.4, respectively. The DDPG controller has the reward function as
\begin{displaymath}
r = 25 -0.5(|s-150|+|v-40|+|u|+|u-u^{'}|),
\end{displaymath}
where 25 is the reward for every successful control and $u^{'}$ is the previous control input (note that this reward function is for learning the underlying controller $\kappa_{2}$, not the Double DQN for adaptation). 

\begin{figure}
    \centering
    \includegraphics[width=\linewidth]{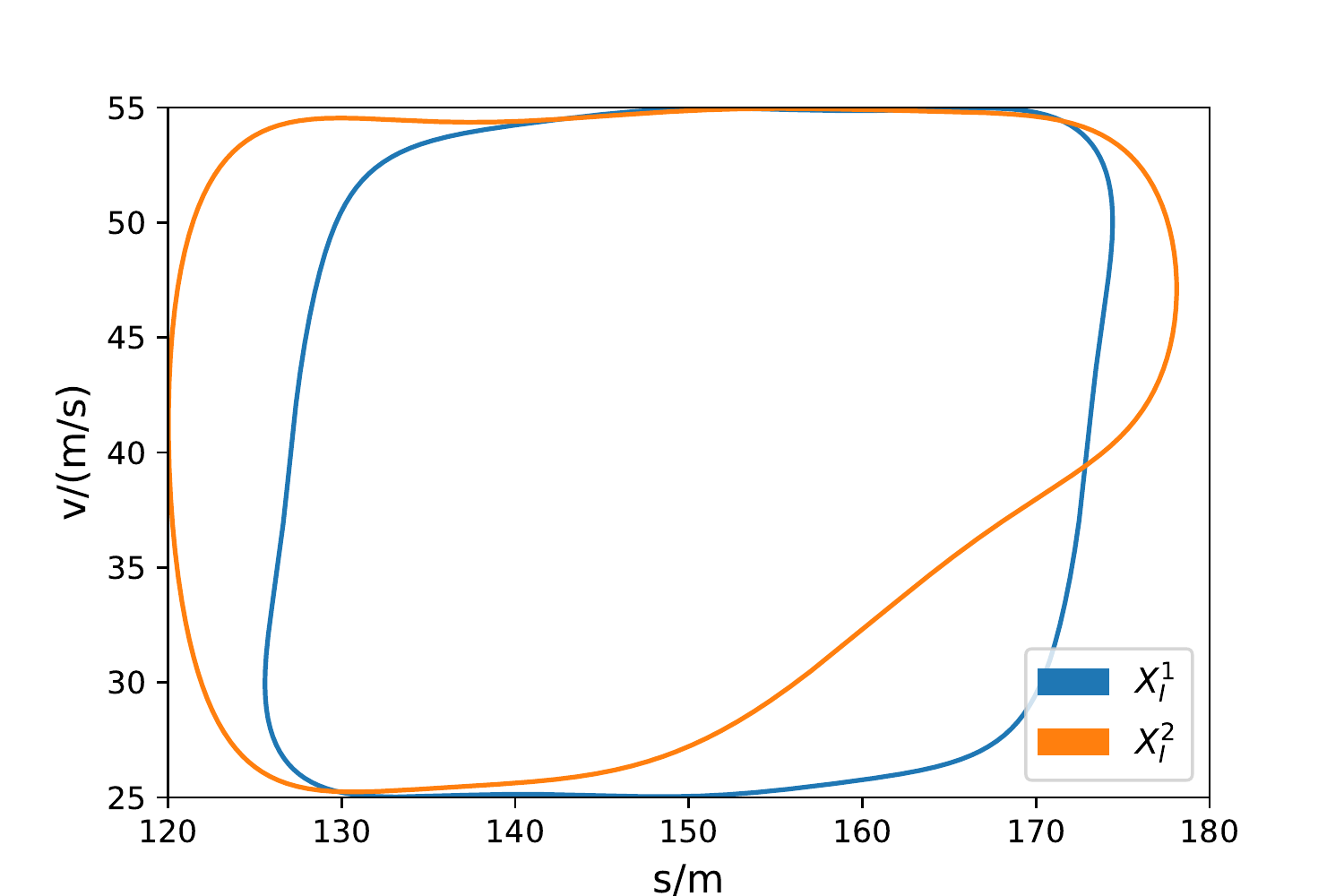}
    \caption{Inner-approximation of the robust invariant sets $X_{I}^{1}, X_{I}^{2}$ for ACC by LQR controller $\kappa_{1}$ and DDPG controller $\kappa_{2}$. Joint safe configuration space is $ X_{I}^{1} \cup X_{I}^{2}$. The controller adaptation learned by Double DQN reduces the overall control energy cost while maintaining the safety.}
    \label{fig:ACC_MI}
\end{figure}

For the LQR controller $\kappa_{1}$, $X_{I}^{1}$ can be directly obtained by SDP.
For the DDPG controller $\kappa_{2}$, Bernstein polynomial approximation via partition is first applied, converting the NNCS into a hybrid polynomial system with bounded disturbance. Then, $X_{I}^{2}$ is obtained for such a hybrid system. $X_{I}^{1}$ and $X_{I}^{2}$ for ACC are shown in Figure~\ref{fig:ACC_MI}.
Then, Double DQN is applied to learn the adaptation strategy. $C$ in Equation (\ref{reward}) is 25,  $\lambda$ is 1, and $R_{pub}$ is -50. The hyper-parameters in Algorithm ~\ref{alg:ddqn} are set as follows: the size of the replay buffer $D$ is 5000, $\gamma$ is 0.99, $C_{0}$ is 100, and the learning rate is 1e-4. 

We consider three baselines: using LQR $\kappa_{1}$ only, using DDPG controller $\kappa_{2}$ only, and random adaptation between the two. We conduct 500 test cases by randomly sampling 500 initial states within $X_{I}^{1} \cup X_{I}^{2}$, and run all the methods from the same initial state for 100 control steps for each case. 

\smallskip
\noindent
\textbf{Comparison among Different Methods:} The comparison of our approach with three baselines are shown in Table~\ref{ACC_result}. Consistent with the results for the Van der Pol's oscillator, our approach achieves the least average energy cost and guarantees $100\%$ safe control rate, outperforming the baselines. Note that in this example, even without the safety guard rule, our approach achieves $100\%$ safe rate (although the safety guard is still needed in practice for guaranteeing safety). 

\begin{table}
  \caption{Comparison of results for the ACC experiment.}
  \label{ACC_result}
  \begin{tabular}{ccccl}
    \toprule
     & Ours & $\kappa_{1}$ only & $ \kappa_{2}$ only & Random\\
    \midrule
    Safe control rate & \textbf{100 \%} & 97.4 \% & 99 \% & 99.6 \% \\
    Energy cost & \textbf{835.7} & 854.8 & 997.5 &  1085.5 \\
  \bottomrule
\end{tabular}
\end{table}

\section{Discussion}\label{sec:discussion}

\textbf{Scale the External Disturbance:} In practice, the system may encounter stronger external disturbance that exceeds the original design expectation. The theoretical robust invariant set of the corresponding system would shrink by some extent in such scenario, and thus safety is no longer guaranteed with the computed invariant. Although, with the inner approximation, the system might still have some buffer to be able to handle such stronger external disturbance. We demonstrate this conjecture in both ACC and oscillator examples by scaling the disturbance to twice and four times of the design assumption. 

\begin{figure}
    \centering
    \includegraphics[width=\linewidth]{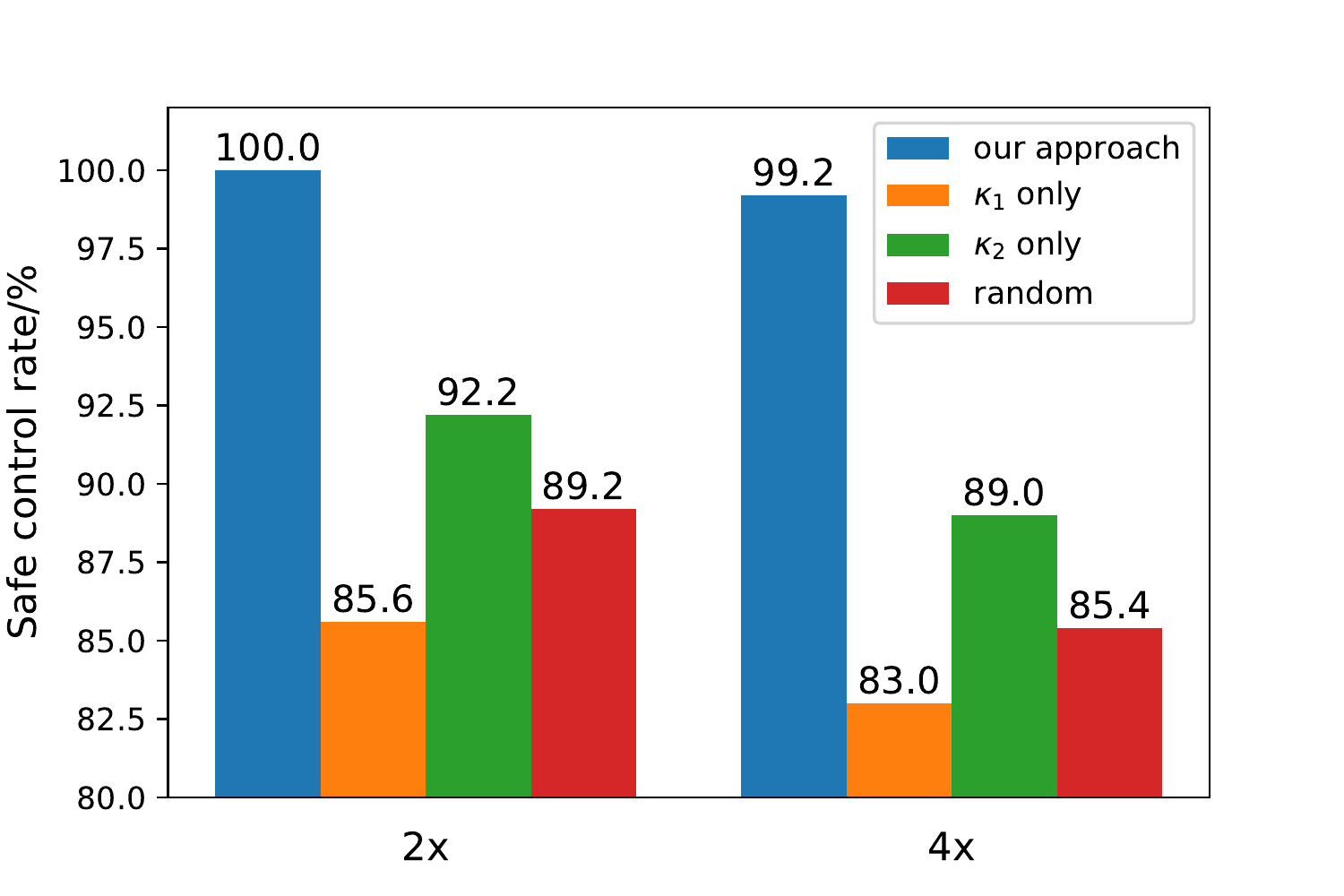}
    \caption{Safe control rate for our approach and the baselines when scaling the external disturbance in the oscillator example by twice (left) and four times (right) of the design assumption.}
    \label{fig:OS_scale}
    \vspace{-15pt}
\end{figure}

\begin{figure}
    \centering
    \includegraphics[width=\linewidth]{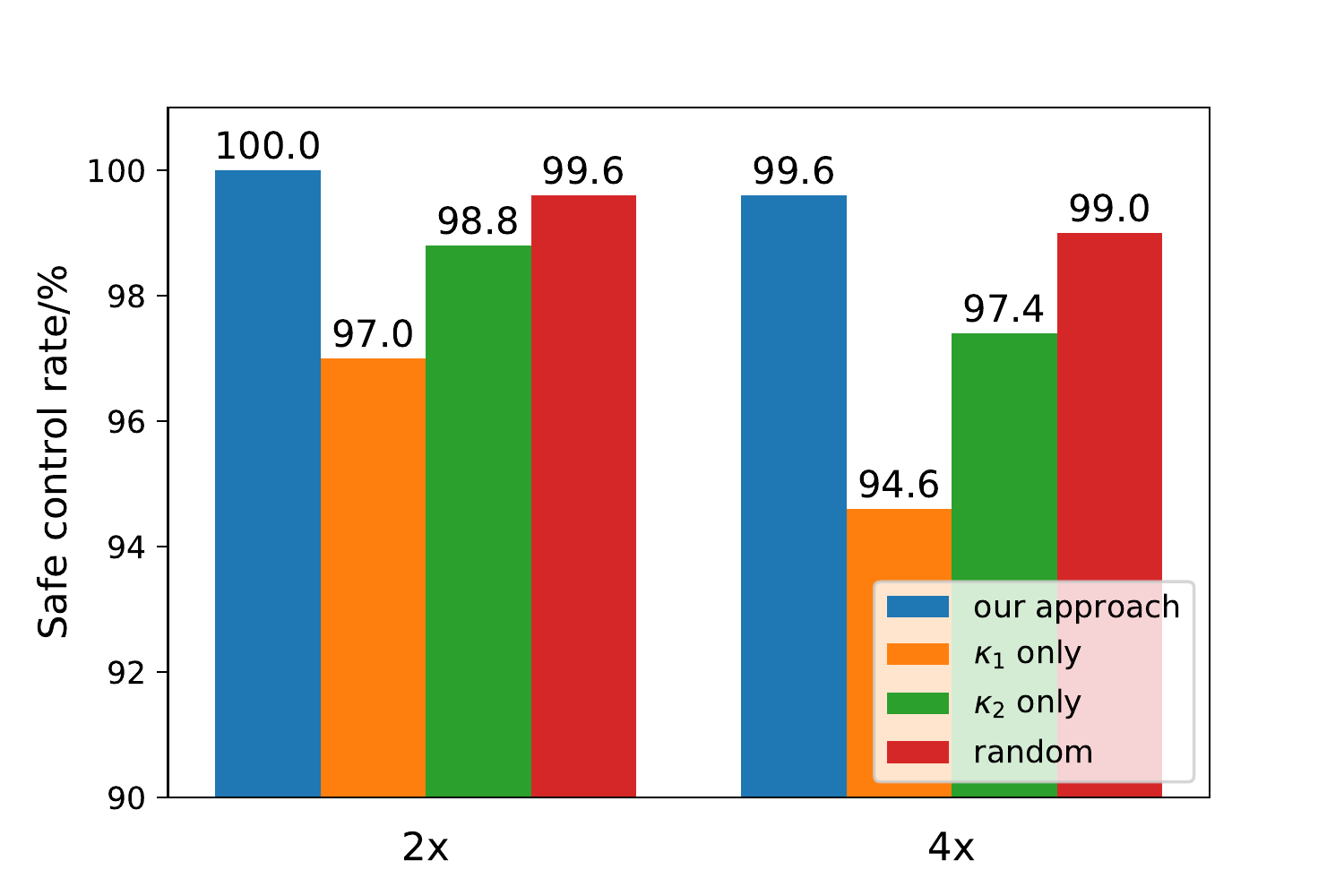}
    \caption{Safe control rate for our approach and the baselines when scaling the external disturbance in the ACC example to twice (left) and fourth times (right) of the design assumption.}
    \label{fig:ACC_scale}
\end{figure}

The results of this study are shown in Figure~\ref{fig:OS_scale}  and~\ref{fig:ACC_scale}. As the disturbance scales, the safe control rates for all methods decrease. However, the safe rate of our approach decreases at a much slower pace than the baselines, showing its robustness to external disturbance (even when the disturbance unexpectedly exceeds the design assumption). Note that the safe rate of our approach is still $100\%$ in the experiments when the disturbance doubles, although this is not always guaranteed.    


\smallskip
\noindent
\textbf{States Outside of the Joint Safe Configuration Space:} There might also be cases in practice where we cannot set the initial state to be within the joint safe configuration space and thus cannot guarantee the system safety. In this study, we conduct experiments to evaluate how our approach performs in such scenario, and how it compares with the baselines. Specifically, we 
train a Double DQN agent with the same reward design on the entire state space $X$, and we do not end a training epoch if the agent chooses an unsafe controller.  

\begin{figure}
    \centering
    \includegraphics[width=\linewidth]{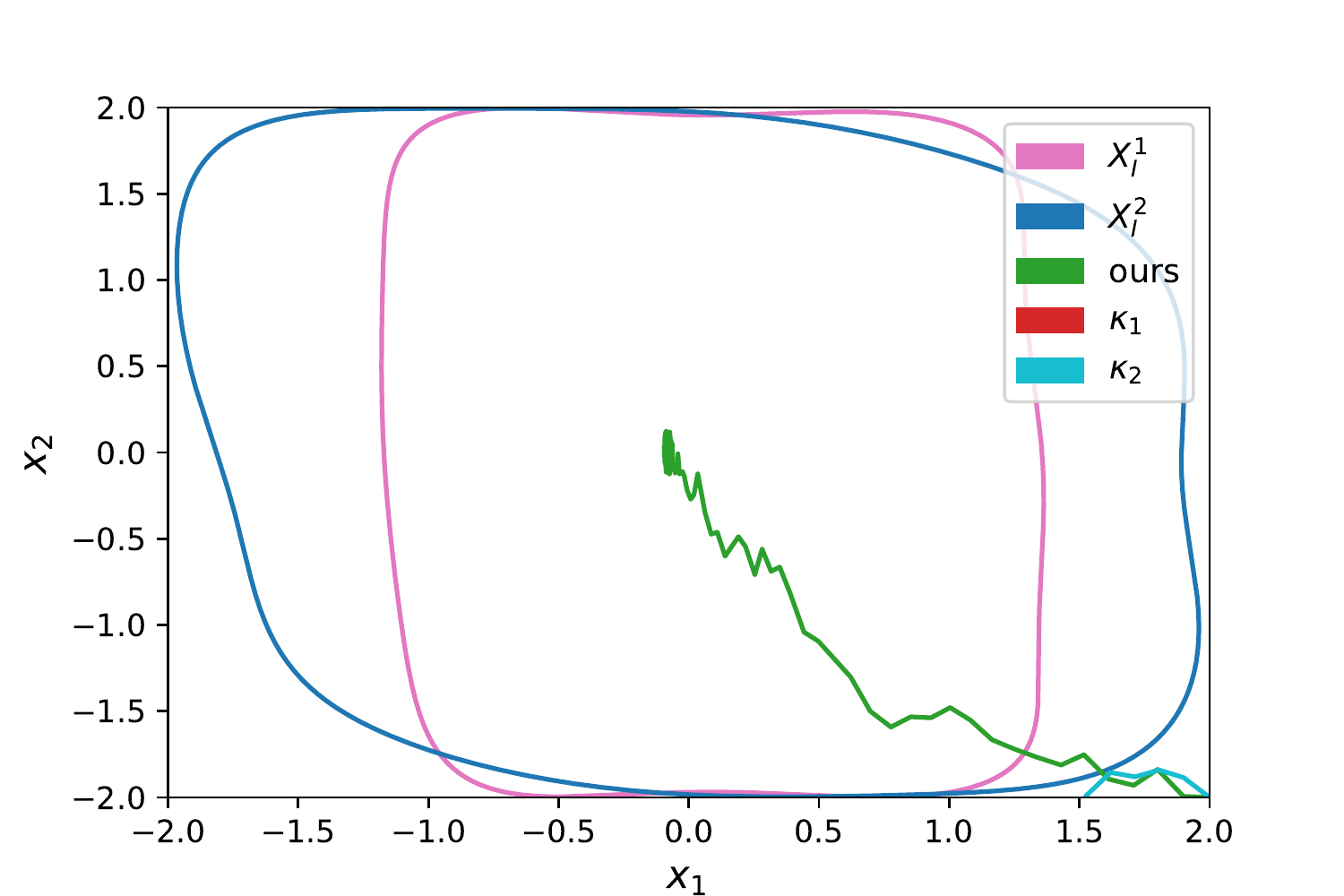}
    \caption{Robust invariant sets and system trajectory under different methods when initial state [2, -2] is outside of the joint safe configuration space for the oscillator example. Our approach is able to pull the state into the joint safe configuration space and maintain system safety. $\kappa_{1}$ fails after one step control (not visible), $\kappa_{2}$ fails after a few steps. (Best viewed in color)}
    \label{fig:OS_dis}
\end{figure}

\begin{figure}
    \centering
    \includegraphics[width=\linewidth]{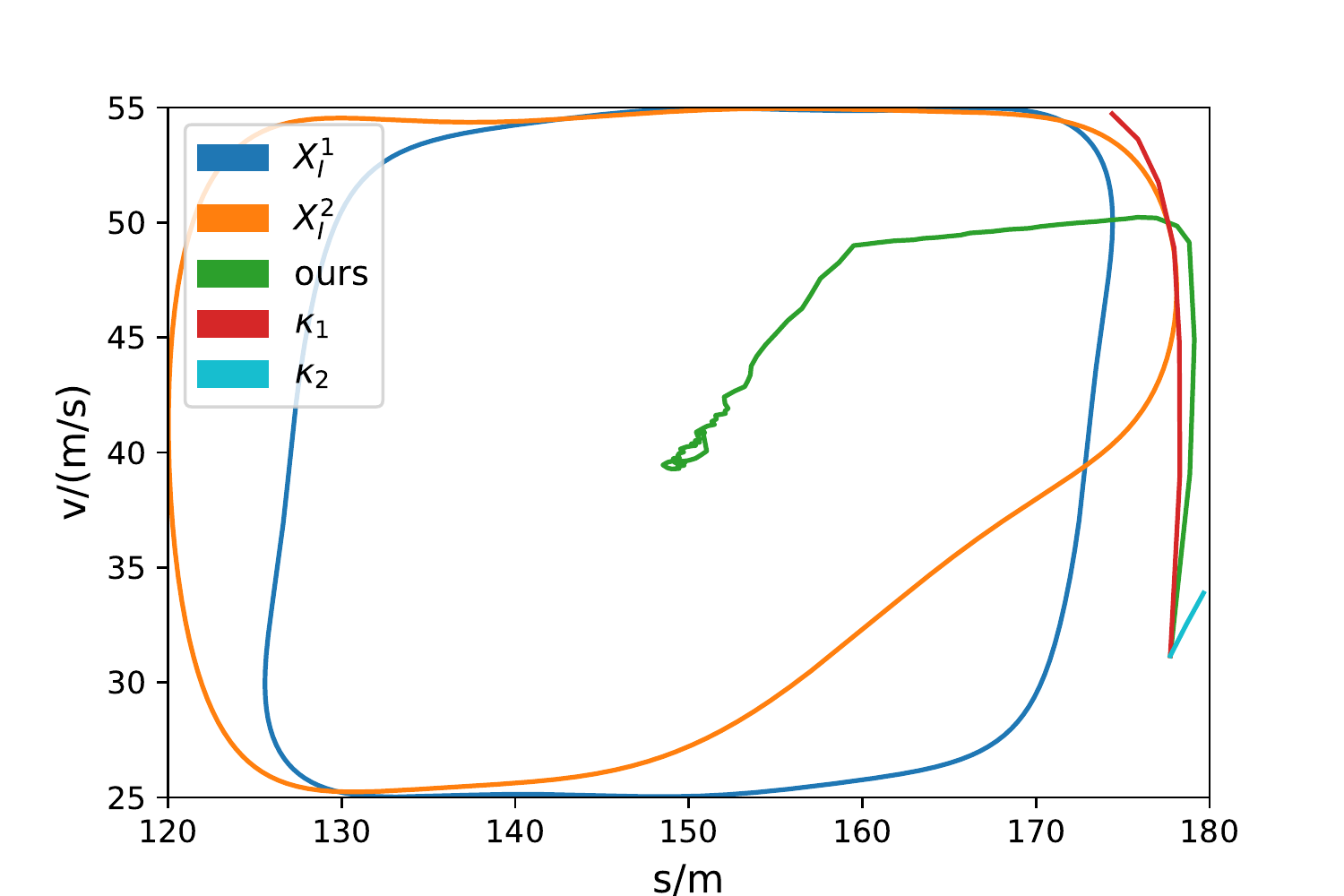}
    \caption{Robust invariant sets and system trajectory under different methods when initial state [177.74, 31.16] is outside of the joint safe configuration space for the ACC example. Our approach can pull the state into the joint safe configuration space and maintain system safety. LQR controller $\kappa_{1}$ and DDPG controller $\kappa_{2}$ both fail after a few steps. (Best viewed in color)} 
    \label{fig:ACC_dis}
\end{figure}

The results for the oscillator example (initial state $x(0) = (-2, 2)$) and the ACC example (initial state $x(0) = (177.74, 31.16)$) are shown in Figure~\ref{fig:OS_dis} and~\ref{fig:ACC_dis}, respectively. We can see that our approach can pull the system state into the joint safe configuration space and then always maintain its safety from that moment, while the baselines with a single controller cannot. This shows that even when the initial state is outside of the joint safe configuration space, our approach may still be able to adapt the system into such space for ensuring system safety.




\smallskip
\noindent
\textbf{Limitation:} It is difficult for our current approach to handle high-dimensional systems. First, it is challenging to accurately approximate neural network controllers with high-dimensional input by Bernstein polynomials. Second, the computation complexity of the robust invariant set increases drastically as the system state dimension increases. Our future work will focus on addressing these issues. 

\section{Conclusions}
\label{sec:conclusion}
We present a controller adaptation approach based on formal methods and machine learning to guarantee system safety and improve energy efficiency for LE-CPSs. In particular, we first compute a joint safe configuration space of the multiple controllers, including neural network ones, with a novel method based on Bernstein polynomial approximation, state partitioning, conversion to hybrid systems, and robust invariant set computation. We then develop a DRL-based method to intelligently switch between controllers for reducing energy consumption while maintaining system safety by keeping its state within the safe space. Experimental results and  analysis on two different case studies demonstrate that our approach significantly outperforms the baselines in both safety and energy efficiency.

\newpage
\bibliographystyle{ACM-Reference-Format}
\bibliography{root}

\end{document}